\documentclass[aps,twocolumn,amsmath,amssymb,reprint,numbers,superscriptaddress,noeprint]{revtex4-1}

\usepackage{makecell}
\usepackage{soul}
\usepackage{geometry}
\usepackage{graphicx}
\usepackage{booktabs}
\usepackage{array}
\usepackage{relsize}
\usepackage{color}
\usepackage{xcolor}
\usepackage{eqnarray}
\usepackage{bm}
\usepackage{textcase}
\usepackage{textcomp}
\usepackage{braket}
\usepackage{setspace}
\usepackage{mathtools}

\usepackage{hyperref}
\hypersetup{
     colorlinks   = true,
     linkcolor    = blue,
     citecolor    = blue,
     urlcolor = blue
}

\definecolor{cola}{rgb}{0.8,0.1,0.1}
\definecolor{colb}{rgb}{0.8,0.3,0}
\definecolor{colc}{rgb}{0.3,0.7,0}
\definecolor{cold}{rgb}{0,0.35,0.75}
\definecolor{cole}{rgb}{0.63, 0.13, 0.94}
\definecolor{colf}{rgb}{0.0, 0.6, 0.0}

\newcommand{\response}[1]{{\color{black} #1}}

\newcommand{\katerc}[1]{{\color{black} #1}}
\def\be{\begin{equation}}
\def\ee{\end{equation}}

\begin{document}

\title[]{Energetic cost of measurements using quantum, coherent, and thermal light}

\author{Xiayu~Linpeng}
\affiliation{Department of Physics, Washington University, St. Louis, Missouri 63130}
\author{L\'ea Bresque}
\affiliation{Universit\'e Grenoble Alpes, CNRS, Grenoble INP, Institut N\'eel, 38000 Grenoble, France}
\author{Maria Maffei}
\affiliation{Universit\'e Grenoble Alpes, CNRS, Grenoble INP, Institut N\'eel, 38000 Grenoble, France}
\author{Andrew N. Jordan}
\affiliation{Institute for Quantum Studies, Chapman University, Orange, California 92866, USA}
\affiliation{Department of Physics and Astronomy, University of Rochester, Rochester, New York 14627, USA}
\author{Alexia Auff\`eves}
\affiliation{Universit\'e Grenoble Alpes, CNRS, Grenoble INP, Institut N\'eel, 38000 Grenoble, France}
\author{Kater W. Murch}
\affiliation{Department of Physics, Washington University, St. Louis, Missouri 63130}

\begin{abstract}
Quantum measurements are basic operations that play a critical role in the study and application of quantum information.
We study how the use of quantum, coherent, and classical thermal states of light in a circuit quantum electrodynamics setup impacts the performance of quantum measurements, by comparing their respective measurement backaction and measurement signal to noise ratio per photon. In the strong dispersive limit, we find that thermal light is capable of performing quantum measurements with comparable efficiency to coherent light, both being outperformed by single-photon light. We then analyze the thermodynamic cost of each measurement scheme. We show that single-photon light shows an advantage in terms of energy cost per information gain, reaching the fundamental thermodynamic cost.


\end{abstract}

\date{\today}

\maketitle

\katerc{Quantum measurements are ubiquitous in quantum mechanics. They raise questions of fundamental nature \cite{ref:Maudlin1995} and are essential operations in emerging quantum technologies \cite{ref:Briegel2009, ref:Deparny2022}. In this view, it is of fundamental and practical importance to understand the cost of measuring in the quantum realm~\cite{ref:Guryanova2020ipm, ref:Deffner2016qwt, ref:Jacobs2012qmf}.} Pioneering contributions have analyzed the thermodynamic cost of measurement over an elementary cycle, as the energy cost of creating correlations between a system and a memory (readout step), followed by the cost of erasing the memory (erasure step)~\cite{ref:Sagawa2009mec}. For a memory with degenerate energy states, the readout step is energetically free, and the overall cost reduces to the erasure cost. Generalizing to non-degenerate energy states, it was shown that the total energy cost of the cycle is always lower bounded by $k_B T_D I$, where $k_B$ is the Boltzmann constant, $T_D$ is the temperature of the memory and $I$ is the mutual information between the measured system and the memory. Comparing the total energy cost of such a cycle to this fundamental bound defines an energetic efficiency for the measurement process.



The circuit quantum electrodynamics (cQED) architectures provide convenient platforms to study the energetic footprint of quantum measurement~\cite{ref:Blais2004cqe,ref:Blais2021cqe}. \response{A microwave cavity plays the role of the memory used to encode a qubit state~\cite{ref:Wallraff2005auv, ref:Walter2017rhf,ref:Jeffrey2014fas}. In this study, we investigate the energy cost of qubit measurements in the strong dispersive limit. Here, the interaction is $H_{\mathrm{int}} = \chi a^\dagger a \sigma_z$ , where $\chi$ is the dispersive shift, $a~(a^\dagger)$ is the annihilation (creation) operator for the cavity, and $\sigma_z$ is the Pauli operator for the qubit. When the dispersive shift $\chi$ is much greater than the cavity dissipation rate $\kappa$,~\cite{ref:Schuster2007rpn}   the qubit state can be distinguished by probing the transmission amplitude of the cavity.} The readout step consists in filling the initially empty cavity with a field, whose final energy depends on the qubit state. This can be performed using coherent, thermal and single-photon light, thereby enabling a direct comparison of the energy cost using different light sources. We examine the measurement backaction in these three scenarios, and quantify their energy cost in terms of emitted cavity photon number. Our analysis reveals that coherent light and thermal light have the same measurement backaction and similar measurement signal to noise ratio (SNR) per photon. We identify an advantage of single-photon light in that it has the lowest energy cost per information gain. In a second step, we theoretically estimate the final meter entropy and subsequent erasure cost assuming there is a Maxwell's demon that can extract the cavity energy. This allows us to quantify the complete energy cost of the measurement-and-erasure cycle and the efficiency of the measurement process. While coherent and thermal light do not operate at maximal efficiency, we show that single photon light saturates the fundamental bound $k_B T_D I$.

\emph{Setup.---}The experimental system comprises a transmon circuit embedded in a 3-dimensional aluminum cavity. The cavity has two ports; a weakly coupled input port and a strongly coupled output port such that intracavity photons predominantly leak out of the output port. The transmon has a qubit  transition frequency of $f^{\mathrm{(q)}}=5.122$~GHz and an anharmonicity of $\alpha/2\pi=-316$~MHz. The cavity frequency depends on the qubit state with $f^{\mathrm{(c)}}_g = 5.6185$~GHz  and $f^{\mathrm{(c)}}_e = 5.6060$~GHz corresponding to the qubit in the ground ($\ket{g}$) and excited ($\ket{e}$) states respectively. When probed at high power, the frequency converges to the bare-cavity frequency $f^{\mathrm{(c)}}_{\text{bare}}=5.6047$~GHz. The cavity is coupled with the qubit in the strong dispersive regime, with a dispersive shift $\chi/2\pi =- 6.3$~MHz and a cavity dissipation rate $\kappa/2\pi = 0.5$~MHz. The qubit has a relaxation time T$_1\simeq9$~\textmu s and a dephasing time T$_2^*\simeq 8$~\textmu s. To perform quantum measurements in this setup, the cavity transmission is probed with different quantum and classical states of light, described below.  

\begin{figure}[!t ]
  \centering
  \includegraphics[width=3.3 in]{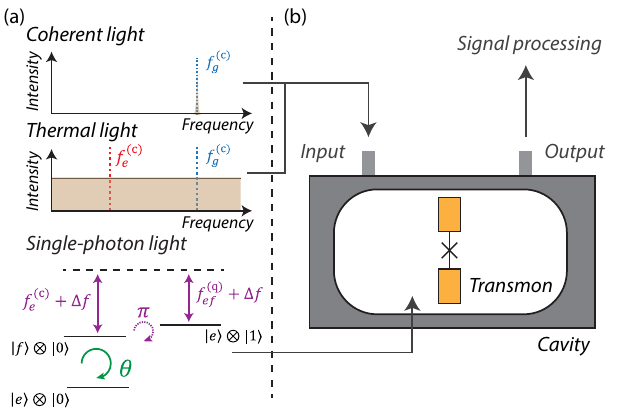}
  \caption{\label{fig:setup} Schematic of the setup. (a) Upper two panels: frequency spectra of the coherent and thermal light. \response{The dashed lines show the frequencies $f^{\mathrm{(c)}}_g$ and $f^{\mathrm{(c)}}_e$, corresponding to cavity resonances with qubit in $\ket{g}$ and $\ket{e}$ states.} Lower panel: Illustration of the effective single-photon light that utilizes a rotation in the $\{e,f\}$ manifold plus two sideband pumps. \response{The green circle arrow represents a rotation between the qubit $\ket{e}$ and $\ket{f}$ states with a rotation angle $\theta$. The two purple arrows represent the two sideband pumps at frequencies $f^{\mathrm{(q)}}_{ef}+\Delta f$ and $f^{\mathrm{(c)}}_{e}+\Delta f$ where $f^{\mathrm{(q)}}_{ef}$ is the frequency of the $\ket{e}\leftrightarrow\ket{f}$ transition and we use a detuning $\Delta f = 0.5$~GHz in the experiment. These two sideband pumps are equivalent to a $\pi$ rotation of the $\ket{f}\otimes\ket{0}\leftrightarrow\ket{e}\otimes\ket{1}$ transition (the dashed purple circle arrow).} (b) Schematic of the cavity-transmon system. \response{The output from the cavity is further demodulated and analyzed to obtain the measurement signal (see~\cite{supplementary} for details of the signal processing).}
  }
\end{figure}

\emph{Coherent light.---}To implement the readout step, we probe the initially empty cavity with a single-frequency microwave tone at frequency $f^{\mathrm{(c)}}_g$ (Fig.~\ref{fig:setup}(a,b)). In the strong dispersive limit ($\chi \gg \kappa$), as the two cavity resonances are well separated, the cavity is excited to a coherent state only if the qubit is in the state $\ket{g}$, changing the quantum states of the qubit and cavity in the following way:
\begin{equation}
\begin{aligned}
\ket{g}\otimes\ket{0} & \rightarrow \ket{g}\otimes\ket{\alpha},  \\
\ket{e}\otimes\ket{0} & \rightarrow \ket{e}\otimes\ket{0}, 
\end{aligned}
\end{equation}
where the state $\ket{i}\otimes\ket{j}$ denotes the qubit ($i=g,e$) and cavity ($j=0,\alpha$) states. \response{Here, $\ket{0}$ is the vacuum state and $\ket{\alpha}$ is the coherent state established by the light where $\alpha$ is a complex value that determines the amplitude and phase of the coherent state. The cavity output is amplified and demodulated to distinguish the qubit states \cite{supplementary}}.



\emph{Thermal light.---}We generate thermal light from a 300~K, 50~$\Omega$ resistor. \response{The Johnson noise from the resistor is filtered, amplified, and attenuated before it is directed to the weakly coupled port of the cavity, resulting in broadband light that uniformly illuminates the $f^{\mathrm{(c)}}_g$ and $f^{\mathrm{(c)}}_e$ resonances of the cavity. A high-pass filter blocks the photons at the qubit transition to prevent decoherence from direct heating of the qubit.} With thermal light, the quantum state of the qubit-cavity system changes as:
\begin{equation}
\begin{aligned}
\ket{g}\otimes\ket{0} & \rightarrow |g\rangle\langle g|\otimes\rho_{\mathrm{th},g},  \\
\ket{e}\otimes\ket{0} & \rightarrow |e\rangle\langle e|\otimes\rho_{\mathrm{th},e}, 
\end{aligned}
\end{equation}
where $\rho_{\mathrm{th},g}$ and $\rho_{\mathrm{th},e}$ correspond to the thermal states generated by the thermal light at frequencies $f^{\mathrm{(c)}}_{g}$ and $f^{\mathrm{(c)}}_{e}$. \response{The cavity output is collected and analyzed using Fourier transform to distinguish the qubit states \cite{supplementary}.} 


\emph{Single-photon light.---}Ideal single-photon illumination would consist of a temporally mode matched single itinerant photon~\cite{ref:Narla2016rcr,ref:Peng2016tds}. Here, we realize an effective single-photon illumination utilizing the $\ket{f}$ state of the transmon to transfer a photon into the cavity. The effective single photon input is realized by first using a resonant rotation on the $\{\ket{e},\ket{f}\}$ manifold by angle $\theta$, mapping the $\ket{e}$ state to a superposition $\cos(\theta/2)\ket{e}+\sin(\theta/2)\ket{f}$. Then, two sideband pumps are applied to yield a coherent rotation between $\ket{f}\otimes\ket{0}$ and $ \ket{e}\otimes\ket{1}$ \cite{ref:Narla2016rcr}, as illustrated in Fig.~\ref{fig:setup}(a). The two sideband pumps used in the experiment are at frequency $f^{\mathrm{(q)}}_{ef}+\Delta f$ and $f^{\mathrm{(c)}}_{e}+\Delta f$, where $f^{\mathrm{(q)}}_{ef} = f^{\mathrm{(q)}}+\alpha/2\pi$ is the frequency difference between the $\ket{e}$ and $\ket{f}$ states of the transmon and $\Delta f$ is a frequency detuning which is set at 0.5~GHz. We set the duration of the sideband pumps so that a $\pi$-pulse is introduced between $\ket{f}\otimes\ket{0}$ and $ \ket{e}\otimes\ket{1}$. Following both rotations, the quantum state of the system changes as:
\begin{equation}
\label{Eq:singlephoton}
\begin{aligned}
\ket{g}\otimes\ket{0} & \rightarrow \ket{g}\otimes\ket{0},  \\
\ket{e}\otimes\ket{0} & \rightarrow \cos{(\theta/2)} \ket{e}\otimes\ket{0} + \sin{(\theta/2)} \ket{e}\otimes\ket{1}. 
\end{aligned}
\end{equation}
The process is identical to single-photon light when ${\theta=\pi}$ and partial single-photon light when ${\theta<\pi}$ with ${n^\mathrm{(c)}=\sin^2(\theta/2)}$ being the average intracavity photon number. Since realizing a single-microwave-photon detector with near-unity efficiency is still a challenging task~\cite{ref:Kono2018qnd,ref:Besse2018ssq, ref:Lescanne2020}, in this work we only experimentally study the backaction of the single-photon source.

\begin{figure}[!t ]
  \centering
  \includegraphics[width=3.3 in]{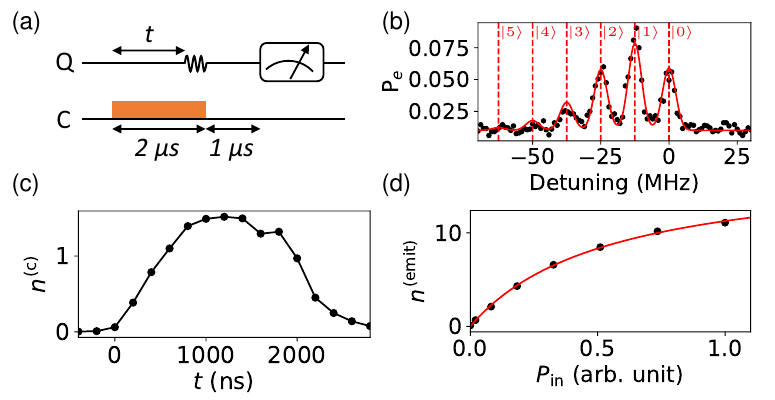}
  \caption{\label{fig:acStark} Determination of the emitted photon number during the measurement. (a) Sequence of the experiment applied to the qubit (Q) and cavity (C). The orange bar represents a square pulse for the measurement light. The projective qubit measurement at the end of the sequence is realized by a high-power readout~\cite{ref:Reed2010hfr}.  \response{(b) A typical qubit spectrum with coherent measurement light at $t=600$~ns exhibits peak splitting with the cavity at different Fock states (dashed lines). The red solid line is a fit using Gaussian peaks following the coherent state distribution.} (c) Typical dynamics of the intracavity photon number $n^{\mathrm{(c)}}$. (d) The emitted photon number $n^{\mathrm{(emit)}}$ as a function of the input power $P_{\mathrm{in}}$ with the maximum power normalized to 1. The red solid line is a fit using an empirical saturation model.
  }
\end{figure}

\begin{figure*}[!t ]
  \centering
  \includegraphics[width=6.3 in]{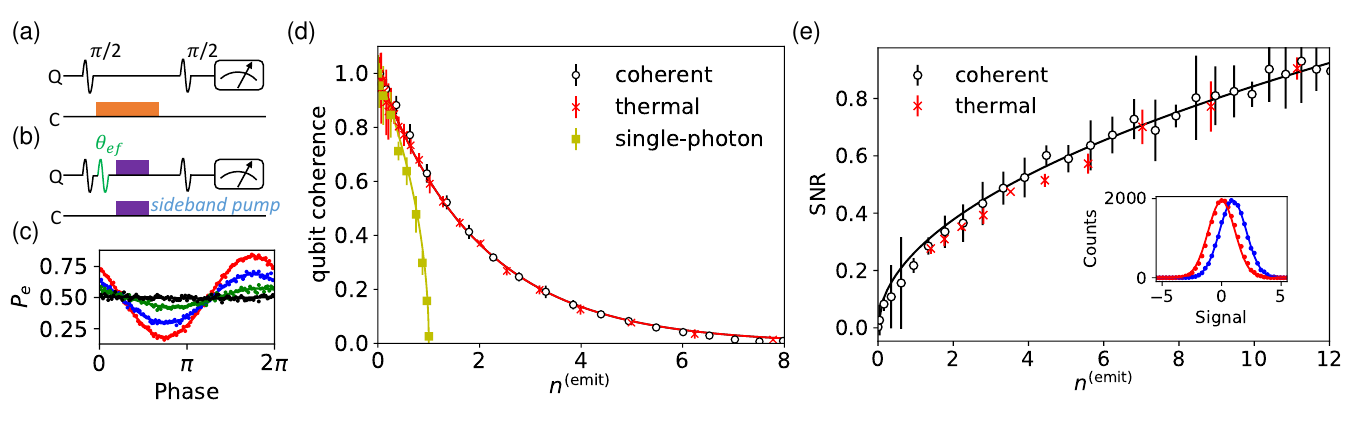}
  \caption{\label{fig:dephasing} Measurement backaction of different light sources. (a) Ramsey experiment sequence for the coherent and thermal light. (b) Ramsey experiment sequence for the single-photon light. $\theta_{ef}$ represents a rotation in the transmon $\{ e,f\}$ manifold with a rotation angle $\theta$. (c) Oscillation of qubit population as a function of the phase on the second $\pi/2$ pulse for coherent light probing. \response{The amplitude of the oscillation is proportional to the qubit coherence. Similar measurements are performed for thermal and single-photon light.} Different colors represent different emitted photon number (red: $n^{\mathrm{(emit)}} = 0$; blue: $n^{\mathrm{(emit)}}=1.0$; green: $n^{\mathrm{(emit)}}=2.8$; black: $n^{\mathrm{(emit)}}=6.5$).  (d) The qubit coherence as a function of the emitted photon number $n^{\mathrm{(emit)}}$ with the maximal qubit coherence normalized to 1. The error bars indicate 2 standard deviations from 5 measurement repetitions. Solid lines are the theoretical prediction~\cite{supplementary}. The red line corresponds to the form $e^{-n^{\mathrm{(emit)}}/2}$ and the yellow line corresponds to the form $\sqrt{1-n^{\mathrm{(emit)}}}$. (e) The measurement SNR for coherent and thermal light as a function of the emitted photon number $n^{\mathrm{(emit)}}$. The error bars indicate 2 standard deviations from 5 measurement repetitions. The black solid line is a fit using a theoretical model for coherent light~\cite{supplementary}. \response{The inset is a typical histogram of the measurement signal using coherent light for qubit at $\ket{g}$ (blue dots) and $\ket{e}$ (red dots) states respectively \cite{supplementary}}. The solid lines are Gaussian fits.
  }
\end{figure*}

\emph{Characterization of the emitted photon number.---}The metric we use to characterize the energy cost of a measurement is the total number of photons emitted by the cavity. For the case of single-photon light, the emitted photon number equals to the intracavity photon number, i.e. $n^{\mathrm{(emit)}}=\sin^2(\theta/2)$. For the coherent and thermal light, as the cavity states are established through quasi-continuous driving, the emitted photon number is determined by the integrated intracavity photon number. We calibrate the intracavity photon number using the ac-Stark effect. In the strong dispersive limit, a splitting can be observed in the qubit spectrum associated with different cavity Fock states \cite{ref:Schuster2005ssd, ref:Schuster2007rpn}.  Figure~\ref{fig:acStark}(a) displays the procedure to obtain the qubit spectrum during the measurement pulse: the light source for the cavity is turned on for a duration of 2~\textmu s and a 200~ns  qubit drive is turned on with varying detuning and varying delay $(t)$ from the start of the cavity pulse. A qubit readout is applied at the end to determine the qubit excitation probability ($P_e$). The resulting qubit spectrum is shown in Fig.~\ref{fig:acStark}(b) for a  coherent light cavity probe. By fitting the qubit spectrum using Gaussian peaks assuming the integrated intensity of each peak following a coherent state distribution, the intracavity photon number is obtained. Figure~\ref{fig:acStark}(c) displays a typical dynamic of the intracavity photon number due to the applied pulse. The total emitted photon number $n^{\mathrm{(emit)}}$ is calculated by $\sum_i n^{\mathrm{(c)}}(t_i) \kappa \Delta t$, where $n^{\mathrm{(c)}}(t_i)$ is the intracavity photon number at time $t_i$, $\kappa$ is the cavity dissipation rate and $\Delta t = t_{i+1}-t_{i} = 200$~ns is the length of the qubit drive. The measured total emitted photon number as a function of the input power is shown in Fig.~\ref{fig:acStark}(d). The saturation of $n^{\mathrm{(emit)}}$ on the input power stems from the cavity nonlinearity~\cite{ref:Reed2010hfr, ref:Boissonneault2010isq}, i.e. the cavity resonances shift at large intracavity photon number. The data displayed in Fig.~\ref{fig:acStark}(b-d) are for coherent light and the corresponding data for thermal light are shown in \cite{supplementary}. 



\emph{Measurement backaction.---}The interaction between the qubit and cavity specifies a natural basis ($\sigma_z$) for measurement. Measurement backaction refers to the reduction of qubit coherences in this basis, and the amount of backaction sets the ultimate limit on extractable information about the qubit \cite{ref:Gambetta2006qpi,ref:Sears2012psn,ref:Hatridge2013qba, ref:Murch2013osq}. We use a Ramsey measurement to characterize this measurement backaction from the three different light sources. The Ramsey experiment consists of two $\pi/2$ pulses with a fixed time delay of 3~\textmu s. For coherent and thermal light, the light source is turned on for 2~\textmu s following the first $\pi/2$ pulse, as shown in Fig.~\ref{fig:dephasing}(a). For single-photon light, the photon is injected after the first $\pi/2$ pulse by using a rotation in the $\{\ket{e},\ket{f}\}$ manifold and then two sideband pumps, as shown in Fig.~\ref{fig:dephasing}(b). The measured qubit state population after the Ramsey sequence oscillates due to the phase change of the second $\pi/2$ pulse, as shown in Fig.~\ref{fig:dephasing}(c). The amplitude of the oscillation is proportional to the remaining qubit coherence. 




The measured results of qubit coherence versus emitted photon number for the three different light sources are shown in Fig.~\ref{fig:dephasing}(d). For single-photon light, as indicated by Eq.~\eqref{Eq:singlephoton}, the qubit coherence is proportional to $\cos(\theta/2)=\sqrt{1-n^{\mathrm{(emit)}}}$. In contrast to coherent and thermal light which cannot achieve projective measurement even at $n^{\mathrm{(emit)}}=8$, for single-photon light, one photon is sufficient to achieve a projective measurement, which indicates an advantage to this quantum light source in the strong dispersive limit. Remarkably, we find the coherent and thermal states have the same strength of measurement backaction. This equivalence is explained by the fact that at the limit of low photon numbers, a thermal field has the same number distribution as a coherent field. Even though the total emitted photon number $n^{\mathrm{(emit)}}$ can be large, as we use quasi-continuous drives for the coherent and thermal light, the driving pulse should be treated as multiple segments and the photon number in each segment is small  (see~\cite{supplementary} for details of the calculation). Note that the measured backaction of the coherent and thermal light differs from what is expected in the weak dispersive regime ($\chi < \kappa$), where at small photon number, the dephasing for thermal light is half of that for coherent light~\cite{ref:Goetz2017psp}.

%
\emph{Measurement signal to noise ratio.---} The backaction characterizes the effectiveness of the premeasurement, i.e. entanglement between the qubit and the cavity. \response{To obtain the information of the qubit state, we now consider the  classical measurement channels.} These classical channels collapse the qubit--cavity entangled states. The performance of the classical measurement channels are characterized by their SNR. \response{For both coherent and thermal light, the histogram of the measurement signals forms a Gaussian distribution and the distribution is different with qubit on different states, as shown in the inset of Fig.~\ref{fig:dephasing}(e) \cite{supplementary}.} 


The measurement SNR, \response{defined as $2|c_g-c_e|/(\sigma_g+\sigma_e)$, where $c_g$ ($c_e$) and $\sigma_g$ ($\sigma_e$) are the center and standard deviation for the two states is shown in Fig.~\ref{fig:dephasing}(e).} The thermal and coherent light yield similar SNR per photon. This equivalence is unique to the strong dispersive limit studied here, and occurs because in this limit, information is encoded in the amplitude, not the phase of the transmitted light. Owing to the broadband nature of the thermal light, it may yield an important advantage in multi-qubit measurements by probing multiple qubits simultaneously without degradation of the SNR.

\begin{figure}[!t]
  \centering
  \includegraphics[width=0.5\textwidth]{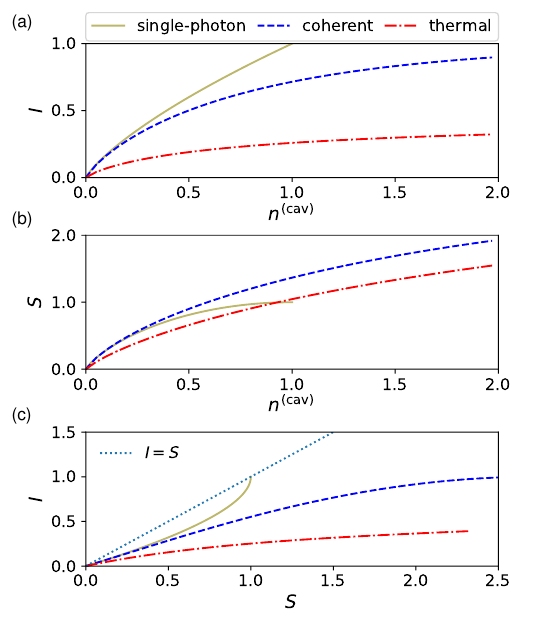}
  \caption{\label{fig:workcost} 
  \response{Fundamental cost of measurement: theoretical insights.} (a) The mutual information $I$ as a function of the cavity photon number $\bar{n}$ for different light sources. (b) The entropy of the three light sources. (c) The ultimate measurement efficiency is given by comparing the information gain ($I$) to the entropy generated in the probe ($S$) with the limit set at ($I=S$), which is only achieved with pure measurement resources such as the single photon state and coherent light.
  }
\end{figure}

\emph{Measurement thermodynamics.}---We now consider the total thermodynamic cost of the measurement for the three light sources. This encompasses the energy cost of the readout and the cost of resetting the cavity back to the vacuum state. The former is the photon energy multiplied by the average number of photons that leave the cavity. In the experiment, the cavity is reset by simply allowing the photons to dissipate into the detector, in which the cavity energy is wasted. While practically simple, this approach is thus highly inefficient from a thermodynamic perspective.

Here we analyze an ideal system where a Maxwell's demon extracts the cavity energy after the readout step. The total energy cost for the whole cycle thus corresponds to the erasure cost of the demon's memory and equals $k_B T_D S$~\cite{ref:elouard2017}, where $S$ is the entropy of the cavity after readout and it is lower bounded by the mutual information $I$ between the cavity and the system. One recovers the fundamental measurement cost when $S = I$~\cite{ref:Sagawa2009mec}. Note that when $T_D$ is at mK-scale, this ideal energy cost is much lower than the work cost needed in our experiments.

In the following analysis, we adopt a simple model where the cavity is treated as a closed system and the Maxwell demon measures the cavity in the Fock state basis, which is experimentally achievable~\cite{ref:Peronnin2020,bretheau2015quantum, ref:Essig2021mpn}. Here, we analyze the results for a qubit at state $(\ket{g}+\ket{e})/\sqrt{2}$ before the readout. 
Figure~\ref{fig:workcost}(a) shows the calculated mutual information as a function of $n^{\mathrm{(cav)}}$ after the cavity is projected in the Fock state basis. An ideal projective measurement corresponds to extracting one bit of information ($I = 1$ bit). For single-photon light, this is achieved at $n^{\mathrm{(cav)}}=1$. For coherent and thermal light, the measurement extracts less information per photon. The entropy $S$ of the cavity after the readout is computed for the three light sources and compared with the mutual information $I$ in Fig.~\ref{fig:workcost}(b-c). At small photon number, all of the three light sources stand below the $S=I$ limit that corresponds to a maximal measurement efficiency. This limit is achieved for a full single-photon readout, i.e. with $n^{\mathrm{(cav)}} = 1$, demonstrating the advantage of this quantum resource.

\emph{Conclusion.---}
We have experimentally characterized the measurement backaction and the corresponding energy cost for coherent, thermal, and single-photon light for a cQED device in the strong dispersive limit. We further analyze the theoretical bound of the work cost. Among the three light sources, we find the single-photon light consumes the minimum amount of energy cost, showing the advantage of this quantum resource. These results could be helpful for the future design of quantum engines in the cQED architecture~\cite{ref:Cottet2017oqm, ref:Quan2006mda}. Additionally, we have demonstrated quantum measurements using thermal light in the strong dispersive limit and have showed that it has similar measurement SNR as the coherent light. This is a cheap resource and easy to implement, with a potential advantage in the measurements of large-scale qubit systems due to its broadband nature.

\vspace{.2in}

\emph{Acknowledgements.}---This work was supported by the John Templeton Foundation Grant No. 61835, \response{the Foundational Questions Institute Fund Grant No. FQXi-IAF19-05, the ANR Research Collaborative Project ``Qu-DICE'' Grant No. ANR-PRC-CES47,} the NSF Grant No. PHY- 1752844 (CAREER) and use of facilities at the Institute of Materials Science and Engineering at Washington University.

\end{document}


\title[]{Supplementary Information: Energetic cost of measurements using quantum, coherent, and thermal light}

\author{Xiayu~Linpeng}
\affiliation{Department of Physics, Washington University, St. Louis, Missouri 63130}
\author{L\'ea Bresque}
\affiliation{Universit\'e Grenoble Alpes, CNRS, Grenoble INP, Institut N\'eel, 38000 Grenoble, France}
\author{Maria Maffei}
\affiliation{Universit\'e Grenoble Alpes, CNRS, Grenoble INP, Institut N\'eel, 38000 Grenoble, France}
\author{Andrew N. Jordan}
\affiliation{Institute for Quantum Studies, Chapman University, Orange, California 92866, USA}
\affiliation{Department of Physics and Astronomy, University of Rochester, Rochester, New York 14627, USA}
\author{Alexia Auff\`eves}
\affiliation{Universit\'e Grenoble Alpes, CNRS, Grenoble INP, Institut N\'eel, 38000 Grenoble, France}
\author{Kater W. Murch}
\affiliation{Department of Physics, Washington University, St. Louis, Missouri 63130}
\date{\today}

\maketitle

In this Supplementary Information, we present additional experimental data and theoretical calculations to support our results in the main text. In Sec.~\ref{sec:photonthermal}, we show the experimental data used to characterize the intracavity photon number for thermal light. \xiayu{Then we present the details on the signal processing of the cavity output in Sec.~\ref{sec:signalProcess}. In Sec.~\ref{sec:fig4theory}, we present the theoretical details to obtain the results in Fig.~4 of the main text.} Section~\ref{sec:backactionTheory} and Sec.~\ref{sec:SNRTheory} present the theories used to calculate the measurement backaction and SNR, respectively. In Sec.~\ref{sec:symmetricCavity}, we extend our discussion to a symmetric cavity probed by thermal light.

\begin{figure}[!b ]
  \centering
  \includegraphics[width = 4.5 in]{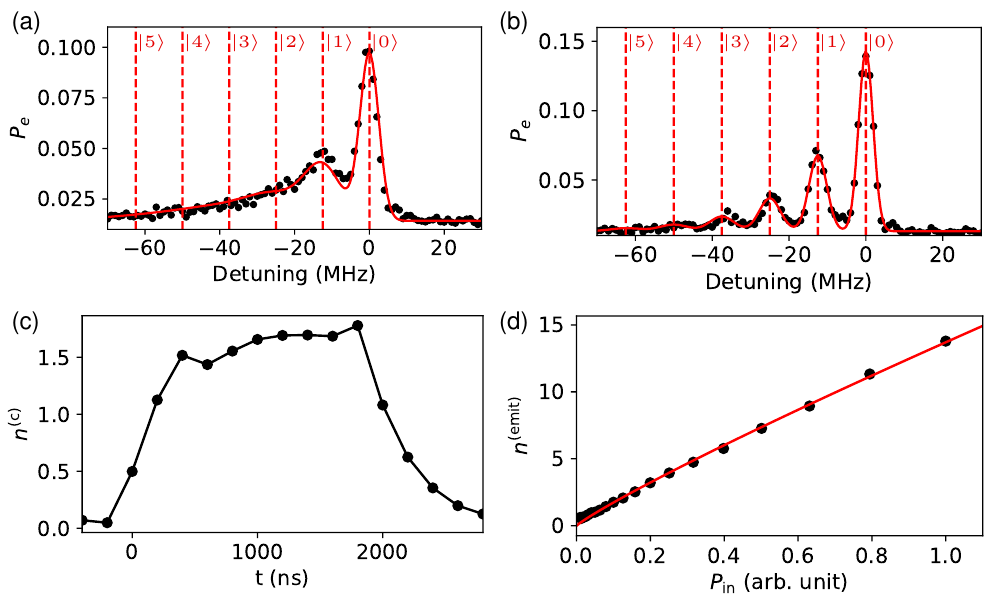}
  \caption{\label{SIfig:acStark} Determination of the emitted photon number for thermal light. (a,b) Typical qubit spectra at time $t<2$~\textmu s and $t>2$~\textmu s. The red dashed lines correspond to qubit frequencies with the cavity at different Fock states. The red solid line is a fit using Gaussian peaks and assuming the integrated intensity of each peak follows the thermal state distribution. (c) Typical dynamics of the intracavity photon number $n^{\mathrm{(c)}}$. (d) The total emitted photon number $n^{\mathrm{(emit)}}$ as a function of the input power $P_{\mathrm{in}}$ with the maximum power normalized to 1. The red solid line is a fit using the saturation model.
  }
\end{figure}

\section{Characterization of intracavity photon number for thermal light}
\label{sec:photonthermal}

The microwave sequence to characterize the intracavity photon number for thermal light is the same as that for coherent light (see Fig.~2(a) in the main text). Figure~\ref{SIfig:acStark}(a,b) show the typical qubit spectra with thermal light at time $t<2$~\textmu s (thermal light on) and $t>2$~\textmu s (thermal light off) respectively. For both spectra, the peaks corresponding to different Fock states can be resolved and the peak intensity follows the thermal state distribution. However, the spectrum at time $t<2$~\textmu s has a much broader linewidth and the linewidth increases significantly with $n$. We attribute this to different cavity induced dephasing with thermal light on and off ~\cite{ref:Sears2012psn}. The linewidth of each peak in the qubit spectrum is $\gamma = \gamma_i + \Gamma(n)$~\cite{ref:Schuster2007rpn}, where $\gamma_i$ is the intrinsic linewidth depending on qubit $T_1$ and $T_2^*$, and $\Gamma(n)$ is the cavity induced dephasing corresponding to cavity at Fock state $\ket{n}$. With the thermal light on, $\Gamma(n) = \kappa (2\bar{n}n+n+\bar{n})$, where $\kappa$ is the cavity dissipation rate and $\bar{n}$ is the intracavity photon number in equilibrium. With the thermal light off, $\Gamma(n) = \kappa n$. The larger $\Gamma(n)$ with thermal light on can well explain the broader spectrum shown in Fig.~\ref{SIfig:acStark}(a).

The intracavity photon number $n^{\mathrm{(c)}}$ is obtained by fitting the qubit spectrum using Gaussian peaks with the integrated intensity of each peak following a thermal state distribution. Figure~\ref{SIfig:acStark}(c) shows typical dynamics of the average intracavity photon number. The total emitted photon number $n^{\mathrm{(emit)}}$ is calculated by $\sum_i 2 n^{\mathrm{(c)}}(t_i) \kappa \Delta t$. The factor of two accounts for the fact that the thermal light can excite both the $f^{\mathrm{(c)}}_e$ and $f^{\mathrm{(c)}}_g$ cavity resonances while the measured intracavity photon number $n^{\mathrm{(c)}}$ is obtained with qubit initially at state $\ket{g}$ only. Figure~\ref{SIfig:acStark}(d) shows the emitted photon number as a function of the input power. The data is fit with the same empirical model $A/(1+B/P_\mathrm{in})$ as the result for coherent light. For coherent light, due to the cavity nonlinearity~\cite{ref:Reed2010hfr, ref:Boissonneault2010isq}, the cavity resonances have frequency shifts at large intracavity photon number, inducing a saturation effect of $n^{\mathrm{(emit)}}$ on the input power $P_{\mathrm{in}}$ (see Fig.~2(d) in the main text). For thermal light, as it is broadband, the frequency shift does not have a significant impact thus $n^{\mathrm{(emit)}}$ almost increases linearly with $P_{\mathrm{in}}$.

\xiayu{
\section{Signal processing of the cavity output}
\label{sec:signalProcess}

\begin{figure}[!b ]
  \centering
  \includegraphics[width = 5 in]{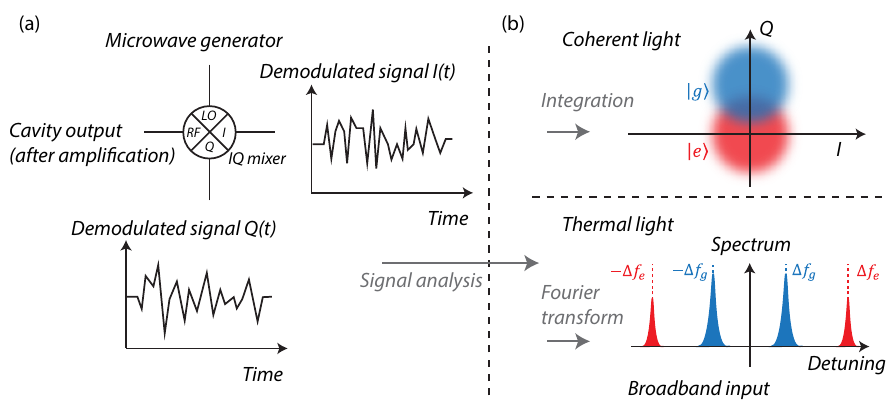}
  \caption{\label{SIfig:signalprocess} \xiayu{Schematic of the signal processing. (a) Demodulation of the cavity output. The cavity output is demodulated using an $I$--$Q$ mixer to obtain the in-phase component $I(t)$ and out-of-phase component $Q(t)$. The frequencies of the microwave generator sent to the LO port of the $I$--$Q$ mixer are $f^{\mathrm{(c)}}_g$ for coherent light and $f^{\mathrm{(c)}}_g+20$~MHz for thermal light. (b) Analysis of the demodulated signals. For coherent light, demodulated signals are integrated over the measurement duration to obtain one data point in the $I$--$Q$ quadrature plane. The distribution of the data points has a Gaussian form and it is different for qubit state $\ket{g}$ and $\ket{e}$. For thermal light, the demodulated signals are analyzed using a Fourier transform. The two peaks near $\pm \Delta f_g = \pm 20$~MHz correspond to qubit state $\ket{g}$ and two peaks near $\pm \Delta f_e = \pm 32.5$~MHz correspond to qubit state $\ket{e}$. }
  }
\end{figure}

Figure~1 of the main text shows the schematic of the setup. At the output of the cavity, the signal is sent for further processing. Figure~\ref{SIfig:signalprocess} shows the details of the processing. First, this cavity output is amplified by a Josephson parametric amplifier operating in phase insensitive mode~\cite{ref:Clerk2010iqn}. Next, following additional stages of amplification, the output is sent to an $I$--$Q$ mixer to obtain the two demodulated signals, i.e. the in-phase component $I(t)$ and out-of-phase component $Q(t)$~\cite{ref:Krantz2019qeg}. The frequency of the demodulation is determined by the frequency of the microwave generator connected to the local-oscillator (LO) port of the $I$--$Q$ mixer. For coherent light, the demodulation frequency is set at the cavity frequency $f^{\mathrm{(c)}}_g$. For thermal light, the frequency is set at $f^{\mathrm{(c)}}_g+20$~MHz. Then, the demodulated signals are collected and analyzed to obtain a measurement signal that can be used to distinguish the qubit state. For coherent light, the demodulated signals $I(t)$ and $Q(t)$ are integrated over the measurement duration. The integrated values form a data point in the $I$--$Q$ quadrature plane. The data points form a Gaussian distribution with a center on the origin for $\ket{e}$ state and with a center in the $Q$ axis for $\ket{g}$ state, as illustrated by the top panel Fig.~\ref{SIfig:signalprocess}(b). The $Q$ value can be used to distinguish the qubit state and we use it as the measurement signal for coherent light. Figure~\ref{SIfig:SNR}(a-b) shows typical experimental data of the distribution in the $I$--$Q$ quadrature plane and the histogram of the measurement signal. For thermal light, we perform a Fourier transform of $I(t) + i Q(t)$  to obtain an amplitude spectrum. The spectrum reveals individual peaks corresponding to the $\ket{g}$ and $\ket{e}$ states of the qubit, as illustrated by the bottom panel of Fig.~\ref{SIfig:signalprocess}(b). The measurement signal is defined by the integrated value of the amplitude spectrum with a weight function, where the weight function is given by the difference between the average spectrum with the qubit initially prepared in state $\ket{g}$ and state $\ket{e}$. The corresponding average amplitude spectra and the histograms of the measurement signal are shown in Fig.~\ref{SIfig:SNR}(c-d). Note that the amplitude spectra in Fig.~\ref{SIfig:SNR}(c) are background subtracted and the background spectra are obtained with the light source off.  

The measurement signals defined here for coherent and thermal light are used to obtain the signal to noise ratio (SNR) in Fig.~3(e) of the main text. The inset of Fig.~3(e) uses the same data as Fig.~\ref{SIfig:signalprocess}(b).



\begin{figure}[!h ]
  \centering
  \includegraphics[width = 3.8 in]{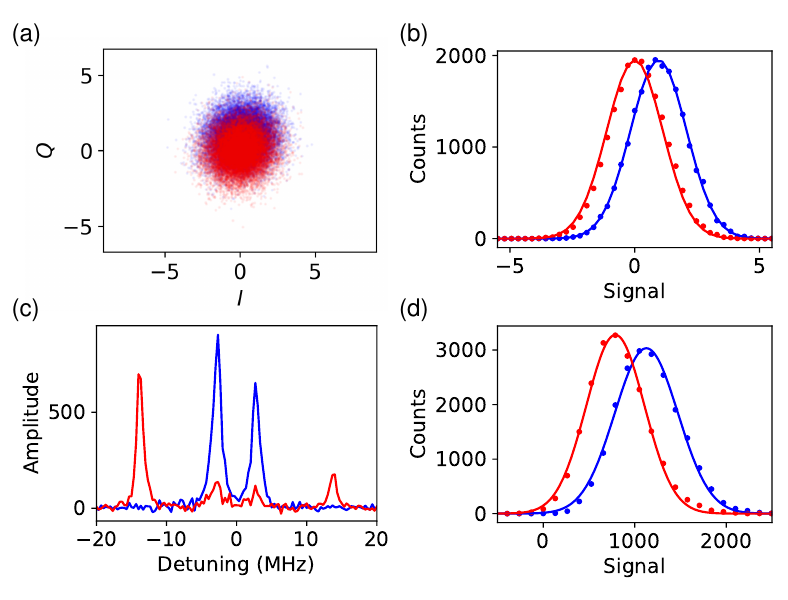}
  \caption{\label{SIfig:SNR} \xiayu{Typical results of the processed output signal and the histogram of measurement signal for coherent and thermal light. For all the subfigures, the red color represents data taken with the qubit in state $\ket{e}$ and the blue color represents data taken with the qubit in state $\ket{g}$. (a, b) Typical distributions in the $I$--$Q$ quadrature plane and the corresponding histograms of the measurement signals for coherent light. Solid lines in the histograms are Gaussian fits. (c, d) Typical amplitude spectrum and the corresponding histograms of the measurement signals for thermal light.}
  }
\end{figure}

}


\section{Theoretical model of measurement backaction} \label{sec:backactionTheory}
The backaction is measured using Ramsey experiments which contains two $\pi/2$ pulses with a fixed time delay. After the first $\pi/2$ pulse, the quantum state of the qubit and the cavity is:
\begin{equation}
    \ket{\psi} =  \ket{+}\otimes \ket{0} =  \frac{1}{\sqrt{2}}\ket{g}\otimes\ket{0}+\frac{1}{\sqrt{2}}\ket{e}\otimes\ket{0} ,
\end{equation}
which defines the initial qubit state $\ket{+}$ and where $\ket{0}$ represents the vacuum state of the cavity. 
\paragraph{Coherent field}
In the case of a coherent input field $\ket{\alpha}$
at the cavity resonance $f_g^{\mathrm{(c)}}$, upon the interaction with a single cell of average number photon $|\alpha|^2 = n^{\mathrm{(emit)}}$,
the joint qubit-field states becomes:
\begin{equation}
    \ket{\psi_\mathrm{coh}} = \frac{1}{\sqrt{2}}\ket{g}\otimes\ket{\alpha}+\frac{1}{\sqrt{2}}\ket{e}\otimes\ket{0}.
\end{equation}
By performing a partial trace over the field, we obtain the qubit's reduced state $\rho_\mathrm{qb}$ and its coherence in the $\{\ket{e},\ket{g}\}$ basis to be:
\begin{align}
\label{eq:coh_deph}
	\rho^\mathrm{(q)}_\mathrm{coh} &= \frac{1}{2}\begin{pmatrix}
			1& e^{-n^{\mathrm{(emit)}}/2} \\
			e^{-n^{\mathrm{(emit)}}/2} & 1 
		\end{pmatrix}
	= e^{-n^{\mathrm{(emit)}}/2}\ket{+}\bra{+} + \frac{1-e^{-n^{\mathrm{(emit)}}/2}}{2}\mathbb{I} \nonumber \\
    |2\rho_{{\mathrm{coh}},{ge}}| &= |\langle \alpha | 0\rangle | = e^{-|\alpha|^2/2} =   e^{-n^{\mathrm{(emit)}}/2}.
\end{align}

\paragraph{Thermal field}
In the case of a thermal field $(\rho_{\mathrm{th},e}\otimes \ket{0} \bra{0}_{g} + \ket{0} \bra{0}_{e} \otimes \rho_{\mathrm{th},g})/2$, where the $g$ and $e$ denotes the modes at frequency $f^{\mathrm{(c)}}_g$ and $f^{\mathrm{(c)}}_e$, the total state and corresponding qubit state writes: 
	\begin{align}
		\rho_\mathrm{th}=&\frac{1}{4} (\ket{e}+\ket{g})(\bra{e}+\bra{g})\otimes(p_{0_e} \ket{0}\bra{0}_e\otimes\ket{0}\bra{0}_g + p_{0_g} \ket{0}\bra{0}_e \otimes \ket{0}\bra{0}_g ) \nonumber \\
		&+  \frac{1}{4}\sum_{n_e \neq 0} p_{n_e} (\ket{e, n_e,0}+\ket{g, 0, 0})(\bra{e,n_e,0}+\bra{g,0,0})+ \frac{1}{4}\sum_{n_g \neq 0} p_{n_g} (\ket{e, 0,0}+\ket{g, 0, n_g})(\bra{e,0,0}+\bra{g,0,n_g})\nonumber \\ 
		\rho_\mathrm{th}^\mathrm{(q)} =& \frac{1}{2}\begin{pmatrix}
			1& p_0 \\
			p_0 & 1  
		\end{pmatrix} = p_0 \ket{+}\bra{+} + \frac{1-p_0}{2}\mathbb{I} 
	\end{align}
	with $p_0 = (p_{0_e} + p_{0_g})/2= (\bra{0}\rho_{\mathrm{th},e}\ket{0} + \bra{0}\rho_{\mathrm{th},g}\ket{0})/2$ and $p_{n_i} = \bra{n}\rho_{\mathrm{th},i}\ket{n}$. 
	Unlike the single photon field, which is created by an almost instantaneous process, the thermal field is built by sending a continuous pulse on the cavity.  This results in the repeated application of $N$ infinitesimal maps each corresponding to $n^{\mathrm{(emit)}}/N$ average photons. 
	Such unital map $\mathcal{M}$ will act on the qubit state $\rho$ such that: 
	\begin{equation}
		\rho \rightarrow p_{0_N} \rho + \frac{1-p_{0_N}}{2}\mathbb{I} 
	\end{equation}
	where $p_{0_N} = \bra{0}\rho_\mathrm{th}^N\ket{0}$ with $\rho_\mathrm{th}^N$ a thermal field with $n^{\mathrm{(emit)}}/2N$ photons on average. 
	Upon $N$ application of this map, the final qubit state becomes: 
	\begin{equation}
		\mathcal{M}^N\rho = (p_{0_N})^N \rho + \frac{1-(p_{0_N})^N}{2}\mathbb{I}. 
	\end{equation}
	The coherence is thus given by $ (p_{0_N})^N = \bra{0}\rho_\mathrm{th}^N\ket{0}^N \approx (1-n^{\mathrm{(emit)}}/2N)^N \approx e^{-n^{\mathrm{(emit)}}/2}$ at first order in $n^{\mathrm{(emit)}}/2N$. 
	Notice that this reasoning could also apply to the coherent case resulting in a coherence for the qubit of $(e^{-n^{\mathrm{(emit)}}/2N})^N = e^{-n^{\mathrm{(emit)}}/2}$ and thus does not change the previous result. 
	This explains why the thermal field and coherent field lead to the same backaction in this experiment.  \\


\paragraph{Single photon field}
For the single-photon light, according to Eq.~(3) of the main text, the quantum state of the qubit and the cavity output is
\begin{equation}
    \ket{\psi_\mathrm{1ph}} = \frac{1}{\sqrt{2}}\ket{g}\otimes\ket{0} + \frac{\cos{(\theta/2)}}{\sqrt{2}} \ket{e}\otimes \ket{0} + \frac{\sin{(\theta/2)}}{\sqrt{2}} \ket{e} \otimes \ket{1}.
\end{equation}
A partial trace over the cavity output states yields the corresponding qubit state and coherence,
\begin{align}
\label{eq:sp_deph}
    \rho^\mathrm{(q)}_\mathrm{1ph} &= \frac{1}{2}\begin{pmatrix}
			1&\cos(\theta/2) \\
			\cos(\theta/2) & 1 
		\end{pmatrix} \nonumber \\
    |2\rho_{{\mathrm{1ph}},ge}| &= |\cos(\theta/2)| = \sqrt{1-n^{\mathrm{(emit)}}}.
\end{align}
The theoretical prediction for the three different light sources is compared with the experimental results in Fig.~3(d) in the main text. The excellent match demonstrates the validity of our model.

Here, we further show that our calculated results for coherent and thermal light are consistent with prior theoretical studies of measurement induced dephasing in cQED~\cite{ref:Gambetta2006qpi, ref:Sears2012psn}. In Ref.~\cite{ref:Gambetta2006qpi}, the authors derived the dephasing rate caused by coherent light in Eq.~(5.12) of the paper:
\begin{equation}
    \Gamma = \frac{(\bar{n}_{+}+\bar{n}_{-})\kappa\chi^2}{\kappa^2/4+\chi^2+\Delta_r^2},
\end{equation}
where $\bar{n}_+$ and $\bar{n}_-$ are the average intracavity photon number with qubit in $\ket{g}$ and $\ket{e}$ states, and $\Delta_r$ is the detuning of the coherent light. In our case, $\Delta_r = -\chi$. In the strong dispersive limit ($\chi \gg \kappa$), we can ignore the $\kappa$ term in the denominator and $\bar{n}_- = 0$. Therefore, $\Gamma = \bar{n}_{+} \kappa /2$. For a pulse duration of time $T$, the qubit coherence is then $|2\rho_{\mathrm{{coh}},{ge}}|=e^{-\bar{n}_{+} \kappa T /2} = e^{-n^{\mathrm{(emit)}} /2} $, which is consistent with the result in Eq.~\eqref{eq:coh_deph}. In Ref.~\cite{ref:Sears2012psn}, the authors derived the dephasing rate caused by thermal light in Eq.~(2) of the paper:
\begin{equation}
    \Gamma = \kappa (2 \bar{n} N + \bar{n} + N),
\end{equation}
where $\Gamma$ is the dephasing rate for the cavity in Fock state $\ket{N}$ and $\bar{n}$ is the average intracavity photon number in equilibrium. In our case, as the qubit is prepared when the cavity is initially in $\ket{0}$ state, the dephasing rate is $\Gamma=\kappa\bar{n}$. For a pulse duration of time $T$, the corresponding qubit coherence is then $|2\rho_{\mathrm{th},{ge}}|=e^{-\bar{n} \kappa T} = e^{-n^{\mathrm{(emit)}} /2} $, which is consistent with our result. Note that the factor of 2 in $\bar{n} \kappa T = n^{\mathrm{(emit)}} /2$ accounts for the fact that the thermal light excites both resonances of the cavity.

\section{Theoretical model of SNR for coherent light} \label{sec:SNRTheory}

With the coherent light, the photon state at the cavity output is $\ket{\alpha}$ with the qubit at state $\ket{g}$, and the photon state is $\ket{0}$ with the qubit at state $\ket{e}$. For the output state $\ket{\alpha}$, the probability to find the state in a coherent state $\ket{\beta}$ is
\begin{equation}
|\langle \beta | \alpha \rangle|^2 = e^{-|\beta-\alpha|^2}. 
\end{equation}
In the $I$--$Q$ quadrature plane, this corresponds to a Gaussian distribution with a standard deviation of $1/\sqrt{2}$ and a center at $\alpha$. The output state $\ket{0}$ corresponds to a Gaussian distribution with the same standard deviation and a center at the $I$--$Q$ origin. Therefore, the SNR for the coherent light is
\begin{equation}
    \mathrm{SNR} = \eta \sqrt{2} \alpha = \eta \sqrt{2 n^{\mathrm{(emit)}}},
\end{equation}
where $\eta$ is the detection efficiency and $\eta < 1$ due to attenuation and noise in the measurement setup. We use this model to fit the SNR data for coherent light in Fig.~4 of the main text. The extracted efficiency $\eta$ is $\sim$20\%. 

\section{Details about figure 4}
\label{sec:fig4theory}
	In Fig.~4 (a) and (b), we plot the field entropy and the mutual information between the field and qubit state for three different input fields: thermal, coherent and single photon fields. 
	This theoretical modelling focuses on the toy model of a cavity dispersively interacting with a qubit. At the beginning, the cavity is in its ground state and the qubit is in $ \frac{\ket{e}+\ket{g}}{\sqrt{2}}$. 
	We plot the Von Neumann entropy of the reduced field state after its interaction with the qubit defined as: 
	\begin{equation}
		S(\rho) = - \text{Tr}(\rho\log(\rho)).
	\end{equation}
	The mutual information $I$ between the qubit and the field is defined in terms of the field entropy $S_\mathrm{f}$, qubit entropy $S_\mathrm{q}$ and total entropy $S_\mathrm{tot}$ in the following way: 
	\begin{equation}
		I = S_\mathrm{q} + S_\mathrm{f} - S_\mathrm{tot}.
	\end{equation}
	For each type of field, the joint qubit-field state is obtained from $\ket{\psi}_\mathrm{1ph}$, $\ket{\psi}_\mathrm{coh}$ and $\rho_\mathrm{th}$ given in the previous section, after the unread measurement of the field in the Fock state basis. 
	The resulting states thus are: 
	\begin{align}
		\rho_\mathrm{1ph} &= \frac{(\ket{e}\cos(\theta/2)+\ket{g})(\bra{e}\cos(\theta/2)+\bra{g})\ket{0}\bra{0}}{2} +\frac{\sin^2(\theta/2)\ket{e}\bra{e}\otimes\ket{1}\bra{1}}{2}\nonumber \\
		\rho_\mathrm{coh} &= \frac{(\ket{e}+\ket{g}e^{-|\alpha|^2/2})(h.c.)\ket{0}\bra{0}}{2} + \sum_{k\neq0}\frac{e^{-|\alpha|^2} \alpha^{2k}\ket{g}\bra{g}\otimes\ket{k}\bra{k}}{2*k!} \nonumber \\
		\rho_\mathrm{th} &= \frac{1}{4} (\ket{e}+\ket{g})(\bra{e}+\bra{g})\otimes(p_{0_e}  + p_{0_g})\ket{0}\bra{0}_e\otimes\ket{0}\bra{0}_g \nonumber \\
		&+  \frac{1}{4}\sum_{n_e \neq 0} p_{n_e} (\ket{e, n_e,0}\bra{e,n_e,0}+\ket{g, 0, 0}\bra{g,0,0})+ \frac{1}{4}\sum_{n_g \neq 0} p_{n_g}(\ket{e, 0,0}\bra{e,0,0}+\ket{g, 0, n_g}\bra{g,0,n_g})\nonumber \\ 
	\end{align}
	In Fig~4(b) of the main text, it is shown that the field entropy is larger in the coherent case than in the thermal case. This might seem a bit surprising and to understand the reason being this difference, we write here the reduced states of these fields: 
	\begin{align}
		\text{Tr}_\mathrm{q}(\rho_\mathrm{coh})&= \ket{0}\bra{0} \frac{1+e^{-|\alpha|^2/2}}{2} + \sum_k \ket{k}\bra{k} \frac{e^{-|\alpha|^2}\alpha^{2k}}{2 k!} \nonumber \\
		\text{Tr}_\mathrm{q}(\rho_\mathrm{th})&= \ket{0}\bra{0} (\frac{p_{0_e}  + p_{0_g}}{2} + \frac{\sum_{n_e\neq 0_e}p_{n_e}}{4}+\frac{\sum_{n_g\neq 0_g}p_{n_g}}{4})+ \frac{\sum_{n_e\neq 0_e}p_{n_e}}{4} \ket{n_e, 0}\bra{n_e, 0} +  \frac{\sum_{n_g\neq 0_g}p_{n_g}}{4} \ket{0,n_g }\bra{0,n_g}.
	\end{align} 
	From these equations we see that, if we restrict ourselves to 0 and 1 photons, the thermal field has three possible states : $\ket{0,0}\bra{0,0} ,\ \ket{0,1}\bra{0,1}$, and  $\ket{1,0}\bra{1,0} $ with respective probabilities $1-n/4,\ n/8$, and $n/8$ whilst the coherent field has only two possible states $\ket{0}\bra{0}$ and $\ket{1}\bra{1}$ with respective probabilities $1-n/2$ and  $n/2$. 
	For small $n$ the coherent state has a smaller probability of being in the state $\ket{0}$, leading to a higher entropy. 

\section{Modeling for a symmetric cavity: SNR using thermal light}
\label{sec:symmetricCavity}

To gain further intuition into thermodynamic concepts in quantum measurement, we extend our theoretical discussion to treat a symmetric cavity coupled to different thermal light sources.


The photons emitted by the cavity are amplified by a phase preserving amplifier corresponding to heterodyne measurement, followed by demodulation, giving quadrature signals $I$ and $Q$, which are noisy, time dependent functions. The time dependent signals $I(t)$ and $Q(t)$ are Fourier transformed, and the power spectrum shows dependence on the qubit state, either the excited or ground state.

Let us begin with a single frequency analysis, where the driving frequency $\omega$ is detuned by $\delta$ from the cavity resonance frequency $\omega_c$.  We can view the resonant cavity as a light scattering problem, so the incoming modes (from left and right) are $a_{in}, b_{in}$, respectively.  The single mode waveguides are scattered with amplitudes ${\cal T}$ for the transmission coefficient, and ${\cal R}$ for the reflection coefficient.  The scattering matrix converts incoming modes into outgoing modes $a_{out}, b_{out}$,
\be
\begin{pmatrix} b_{out} \\ a_{out} \end{pmatrix}
= \begin{pmatrix} {\cal T}_\pm & {\cal R}_\pm \\
{\cal R}_\pm & {\cal T}_\pm \end{pmatrix} 
\begin{pmatrix} a_{in} \\ b_{in} \end{pmatrix},
\ee
where the $\pm$ refers to the qubit being in either the ground or excited state, pulling the cavity frequency.
Here, we assume a spatially symmetric scattering cavity, so no additional phases are acquired.  The transmission coefficient, for a simple single mode cavity, takes the form,
\be
{\cal T}_\pm = \frac{-\kappa}{\kappa + i (\delta \pm \chi/2)}.
\ee
Let us stress these equations apply in the frequency space, where $\delta = \omega_c - \omega$.
The input state of light comes from a thermal source on the left side with temperature $T_H$.  Let us generalize this analysis by putting a thermal source on the right side with temperature $T_C$.  Thermal states are characterized by the density matrix
\be
\rho = \int \frac{d^2\alpha}{\pi} P(\alpha) |\alpha\ra \la \alpha|,        
\ee
where the Q-distribution is given by
\be
P(\alpha) = \frac{1}{\bar n} \exp\left({-\frac{|\alpha|^2}{\bar n}}\right),
\ee
where the function $\bar n$ is mean occupation number of photons, specified by the Bose-Einstein function
\be
{\bar n} = \frac{1}{\exp\left( \frac{\hbar \omega}{k_B T}\right) -1}.
\ee
The light is in a mixed state, independently occupied for a continuous range of frequencies.

Phase preserving amplification of the outgoing mode gives rise to two results, which may be viewed as the real and imaginary parts of a complex number $\beta$, which is a random number drawn from the distribution
\be
P(\beta) = \la \beta | \rho | \beta \ra,
\ee
where $|\beta\ra$ is a coherent state and $\rho$ is the density matrix of the light.  If we amplify a thermal state with this technique, we obtain the distribution
\be
P(\beta) = \frac{1}{\pi {\bar n}} \int d^2 \alpha \exp\left({-\frac{|\alpha|^2}{\bar n}} - |\alpha - \beta|^2\right) = \frac{1}{{\bar n}+1} \exp\left({-\frac{|\beta|^2}{{\bar n}+1}}\right),
\ee
where we have used the inner product $|\la \alpha | \beta \ra|^2 = \exp(-|\alpha - \beta|^2)$.  Thus, we see the amplified state is a photon added thermal state, which is the minimum amount of noise a quantum limited amplified adds to the $I,Q$ quadratures.  
From this distribution, we see that 
\be
\la {\rm Re}\beta \ra = \la {\rm Im}\beta \ra = 0, \qquad
\la ({\rm Re}\beta)^2 \ra = \la ({\rm Im}\beta)^2 \ra =  \frac{{\bar n} + 1}{2}.
\ee

The quadrature signals come from a time-sequence of such measurements $\{ \beta_j \}$, where $j = 1, \ldots, N$, for a quasi-continuous function $I(t)$ and $Q(t)$. 
We predict, therefore, that the time dependent quadrature signals $I(t)$, $Q(t)$ have zero average, so there is no information.  However, we can also look at the Fourier transformed signal over a period of time $T$, and look at the power in the Fourier transform,
\be
\frac{1}{T} \la|{\tilde I}(\omega)|^2 \ra = \frac{1}{T} \int_{-T/2}^{T/2} dt_1 dt_2 e^{i \omega(t_1-t_2)} \la I(t_1) I(t_2)\ra = \int d\tau e^{i \omega \tau} \la I(t) I(t+\tau)\ra.
\ee
If we had only the thermal light, the spectral weight would be centered at zero frequency. However, the spectral filtering of the cavity gives a nontrivial spectral weight to the output.

Amplifying the mode $b_{out} = {\cal T}_{\pm}(\omega) a_{in} + {\cal R}_{\pm} b_{in}$, we can calculate the spectral power because these modes are already expressed in the frequency space.  The spectral power is given by $S(\omega) = \la \beta^\ast \beta \ra$.  The light from the left and right sides is uncorrelated, so we have the power spectral, given the qubit is in the excited or ground state, is given by 
\be
S_{\pm}(\omega) = |{\cal T}_{\pm}(\omega)|^2 \left( \frac{{\bar n}_H(\omega) + 1}{2}\right) +  |{\cal R}_{\pm}(\omega)|^2 \left( \frac{{\bar n}_C(\omega) + 1}{2}\right) ,
\ee
where we recall that the light from the left side is taken from a thermal blackbody source of temperature $T_H$ and light from the right side is taken from a blackbody source of temperature $T_C$.
Thus, we see that the signal is in the noise.  It is important to notice that if the temperatures of the two thermal reservoirs are the same, $T_H = T_C$, then the occupations are the same, ${\bar n}_H = {\bar n}_C$, and because the reflection plus transmission coefficients must be 1, $|{\cal T}_{\pm}|^2 + |{\cal R}_{\pm}|^2 =1 $, then the power spectrum is independent of the qubit state. This indicates thermal resources in global thermal equilibrium cannot be used as a detector.

We define the frequency-resolved qubit signal as the difference $S(\omega) = S_+ - S_-$, which is then given by
\be
S = {\bar n}_H ( |{\cal T}_{+}|^2 - |{\cal T}_{-}|^2) +{\bar n}_C ( |{\cal R}_{+}|^2 - |{\cal R}_{-}|^2).
\ee
Here we assume the dependence of the mean occupations on frequency is weak compared to the sharp resonance feature.  For the simple single resonance model, the transmission difference takes the form,
\be
|{\cal T}_{+}|^2 - |{\cal T}_{-}|^2 = \frac{1}{1 + (\delta +\chi/2)^2/\kappa^2} - \frac{1}{1 + (\delta -\chi/2)^2/\kappa^2)}
= \frac{-2 \chi \delta}{(1+(\delta +\chi/2)^2/\kappa^2)(1+(\delta -\chi/2)^2/\kappa^2)},
\ee
with a similar expression for the reflection coefficient difference.

To find the noise in this signal, we must consider the noise of the noise, or the variance of $(b^\dagger_{out} b_{out})^2$.  In the amplified signal, this is related to finding the 4$^\mathrm{th}$ moment of $\beta$, or $\la |\beta|^4 \ra$ in the photon added thermal state.  A straightforward calculation indicates the variance is $({\bar n}+1)^2$. We drop any dependence on the qubit state for the noise calculation as an approximation to find the power signal to noise ratio SNR$_\mathrm{p}$ is given by
\be
\mathrm{SNR}_\mathrm{p} = \frac{S \sqrt{T}}{\sqrt{({\bar n}_H + 1)^2 |{\cal T}|^2 + ({\bar n}_C + 1)^2 |{\cal R}|^2}},
\ee
where $T$ is the duration of the signal.

In the experiment, a frequency integrated version is considered, where to get better signal, the signal is integrated over the linewidth of the resonance before it is differenced.  We can do this integral by assuming that $\chi \gg \kappa$ so the resonances are well separated.  Then, the integral of the transmission coefficient is approximated as,
\be
\int_{\kappa \mp \chi/2}^{-\kappa \mp \chi/2} d \delta \frac{1}{1 + (\delta \pm \chi/2)^2/\kappa^2} \approx \pi \kappa,
\ee
where we extended the limits of the integral to infinity as an approximation. We then define the new signal ${\cal S}$ as the difference of the frequency integrated peaks to find,
\be
{\cal S} = \pi \kappa ({\bar n}_H - {\bar n}_C).
\ee
We can then write the power signal to noise ratio $\overline{\mathrm{SNR}}_\mathrm{p}$ of the frequency integrated signal as 
\be
\overline{\mathrm{SNR}}_\mathrm{p} = \sqrt{\pi \kappa T} \frac{{\bar n}_H - {\bar n}_C}{\sqrt{({\bar n}_H +1)^2 + ({\bar n}_C + 1)^2}},
\ee
where $T$ is the integration time of the experiment.

We see even in the case when the cold bath is at zero temperature, there is vacuum noise from the amplification that enters into the noise term.

%